Article

# Phosphate Vibrations Probe Electric Fields in Hydrated Biomolecules: Spectroscopy, Dynamics, and Interactions

*Published as part of The Journal of Physical Chemistry virtual special issue "Yoshitaka Tanimura Festschrift".*

Thomas Elsaesser,* Jakob Schauss, Achintya Kundu, and Benjamin P. Fingerhut

Read Online

ACCESS | 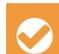 Metrics & More | 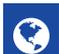 Article Recommendations |

**ABSTRACT:** Electric interactions have a strong impact on the structure and dynamics of biomolecules in their native water environment. Given the variety of water arrangements in hydration shells and the femto- to subnanosecond time range of structural fluctuations, there is a strong quest for sensitive noninvasive probes of local electric fields. The stretching vibrations of phosphate groups, in particular the asymmetric $(PO_2)^-$ stretching vibration $\nu_{AS}(PO_2)^-$, allow for a quantitative mapping of dynamic electric fields in aqueous environments via a field-induced redshift of their transition frequencies and concomitant changes of vibrational line shapes. We present a systematic study of $\nu_{AS}(PO_2)^-$ excitations in molecular systems of increasing complexity, including dimethyl phosphate (DMP), short DNA and RNA duplex structures, and transfer RNA (tRNA) in water. A combination of linear infrared absorption, two-dimensional infrared (2D-IR) spectroscopy, and molecular dynamics (MD) simulations gives quantitative insight in electric-field tuning rates of vibrational frequencies, electric field and fluctuation amplitudes, and molecular interaction geometries. Beyond neat water environments, the formation of contact ion pairs of phosphate groups with $Mg^{2+}$ ions is demonstrated via frequency upshifts of the $\nu_{AS}(PO_2)^-$ vibration, resulting in a distinct vibrational band. The frequency positions of contact geometries are determined by an interplay of attractive electric and repulsive exchange interactions.

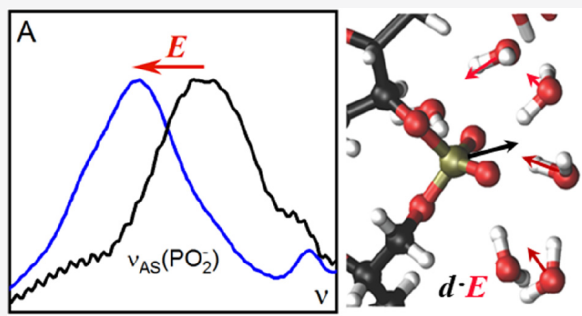

## 1. INTRODUCTION

DNA and RNA consist of sequences of nucleotides which are arranged in a variety of single- or double-stranded geometries, including double-helical structures.[1] Under native conditions, they are embedded in a hydration shell and subject to strong electric fields which originate from the dipole moment of water molecules and from solvated ions.[2−4] Vice versa, the charged phosphate groups of the nucleotides influence the structure and dynamics of the aqueous environment.[2,5] In equilibrium, water molecules and ions are part of a spatial arrangement that minimizes the overall electrostatic energy and at the same time stabilizes the biomolecular structure.[6−8] Thermally excited intra- and intermolecular molecular motions result in structural fluctuations of this many-body ensemble on femto- to subnanosecond time scales. Correspondingly, electric forces between the different molecular entities are not static but display a broad frequency spectrum which extends from a few gigahertz to some 20 THz.

The strength, spatial range, and impact of electric interactions on structure, dynamics, and function of nucleic acids are understood insufficiently. Different theoretical approaches such as Poisson−Boltzmann (PB) theory[9−11] and microscopic molecular dynamics treatments[12−15] have led to conflicting results. PB theory treats the water shell as a dielectric continuum with a static dielectric constant and has been applied, for example, to predict radial distributions of ions around DNA and RNA. In contrast, molecular dynamics simulations include a limited number of water molecules and ions at the molecular level and give insight in local interaction geometries and dynamics of electric forces.

Numerous experiments have focused on time-averaged structural and electric properties from which local electric fields and interaction strengths can be inferred indirectly at best.[3,4,16,17] A more direct approach is based on local molecular probes with field-dependent static or transient spectroscopic properties.[18−20] Both units of the biomolecular structure and/or additional electronic or vibrational chromophores at specific locations have been used as probes. Key issues of electric field probing are the location of the probe in the molecular ensemble and the potential distortion of molecular structure and dynamics by the presence of the probe itself. The least invasive probes are electronic or vibrational excitations of groups in the genuine molecular structure which are located at particular interaction sites.

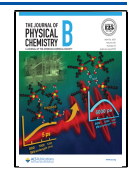

The interface of DNA and RNA with the water shell represents a most relevant region for mapping electric interactions. Recently, we have introduced vibrations of the sugar–phosphate backbone as noninvasive probes at this interface and studied their behavior by stationary vibrational spectroscopy, ultrafast 2D-IR methods, and theoretical calculations.[21,22] In the manifold of backbone vibrations, the stretching modes of the charged and highly polarizable $(PO_2)^-$ groups stand out with a particular sensitivity to external electric fields.[23–26] The two oxygen atoms of $(PO_2)^-$ interact with the water shell directly.[5] As a result, the symmetric and asymmetric stretching vibrations of the $(PO_2)^-$ groups display vibrational frequencies and line shapes markedly influenced by interactions with water molecules and ions. In this way, the electric properties of hydrated DNA and RNA become accessible to vibrational spectroscopy.

The application of $(PO_2)^-$ vibrations as electric field probes in hydrated biomolecules of increasing structural complexity requires a proper quantitative benchmarking of their interactions and a calibration of spectroscopic properties, for example, vibrational transition frequencies, as a function of interaction strength. To achieve this, experimental and theoretical work on small model systems such as dimethylphosphate (DMP) and tailored short DNA and RNA sequences, both single- and double-stranded, are required. In this article, we report a systematic study of phosphate electric field probes, covering a range from DMP in neat water to transfer RNA (tRNA), an extended folded RNA structure with some 80 nucleotides in an aqueous environment. In parallel to the impact of water molecules, we study the interaction of excess magnesium ions $(Mg^{2+})$ with the different systems. Linear stationary infrared spectroscopy and nonlinear 2D-IR techniques are combined with in-depth theoretical calculations of interaction patterns and MD simulations. The results allow for quantifying the strength and dynamics of electric fields and give insight in hydration patterns at the molecular level.

## 2. MATERIALS AND METHODS

Commercial samples of DMP (Sigma-Aldrich), DNA duplexes (Thermo Scientific), RNA duplexes (Integrated DNA Technologies), and tRNA from *Escherichia coli* (E.c.) (Sigma-Aldrich) were dissolved in ultrapure water (Rotipuran Ultra, Roth). The double-stranded DNA oligomers form a B-double helix containing 23 alternating adenine-thymine (A-T) base pairs in a Watson–Crick pairing geometry ($Na^+$ counterions) while the RNA duplexes with 23 alternating adenine-uracil (A-U) pairs exist in an A-double helix geometry ($K^+$ counterions). The tRNA sample represents a mixture of structures with different codon and acceptor stem units. The sample concentrations are given in the respective figure captions. To remove residual concentrations of $Mg^{2+}$ ions, the purely aqueous E.c. tRNA samples were dialyzed, applying the procedures of ref 27.

To the aqueous solutions, excess concentrations of $MgCl_2$ were added, resulting in a concentration $c(Mg^{2+})$ of magnesium ions. In the following, we use the quantity $R = c(Mg^{2+})/c(s)$ with s = DMP, RNA, tRNA, that is, the ratio of the $Mg^{2+}$ concentration to that of DMP, RNA, or tRNA, for distinguishing different levels of $Mg^{2+}$ excess. For determining the fraction of $Mg^{2+}$ ions interacting with RNA duplexes and tRNA in the samples with $Mg^{2+}$ excess, the fluorescence titration method of refs 27–29 was used.

Linear infrared absorption spectra were recorded with a Fourier transform infrared spectrometer (Bruker Vertex 80, spectral resolution 2 $cm^{-1}$). The 2D-IR spectra were measured with a three-pulse photon-echo setup in which 3 femtosecond mid-infrared pulses interact with the sample and a fourth synchronized pulse serves for heterodyning the nonlinear signal diffracted from the sample. The pulses were centered at frequencies between 1200 and 1300 $cm^{-1}$ with a spectral bandwidth of up to 150 $cm^{-1}$. The pulse duration was between 80 and 120 fs, the pulse energy up to 2 $\mu J$. Details of the 2D-IR setup and of pulse generation have been given in ref 21.

Molecular dynamics (MD) simulations follow the procedure described in refs 22 and 25. In brief, MD simulations of alternating $(AT)_{23}$ duplex DNA and $(AU)_{23}$ duplex RNA were performed with AMBER 18[30] employing the *ff99bsc1*[31] force field and used $\chi_{OL3}$ corrections[32] in the case of RNA. The TIP4P-FB[33] water model together with the 12-6-4 Lennard-Jones-type nonbonded model[34,35] for $Na^+$ and $Mg^{2+}$ (TIP4P-FB/12-6-4 LJ) was used. Geometries for the evaluation of electric fields for phenylalanine tRNA ($tRNA^{Phe}$) were taken from ref 29 and used the TIP3P water model together with the nonbonded ion model[34,35] for $Na^+$ and $Mg^{2+}$. A number of 9 $Mg^{2+}$ ions and 56 $Na^+$ ions were included. Initial model structures of DNA and RNA were generated with the nucleic acid builder (NAB) as canonical double helices of B- and A-form with a Watson–Crick base pairing, respectively, placed in a truncated octahedral solvation box with a 10.0 Å buffer region, and 44 $Na^+$ ions were added for charge neutrality. Further, 20 $Mg^{2+}$ and 40 $Cl^-$ ions were added. The number of $Mg^{2+}$ ions for $(AT)_{23}$ duplex DNA and $(AU)_{23}$ duplex RNA correspond to a concentration of 0.067 and 0.081 M, respectively. $Na^+$ concentrations are 0.148 and 0.179 M, close to a 0.15 M physiological concentration. Simulations were performed with the PMEMD program and the GPU accelerated PMEMD.CUDA program (Tesla V100).[36,37]

Electric fields were calculated at the bisector midpoint of the $(PO_2)^-$ groups followed by projection on the O1=P=O2 bisector axis which gives the projected field $E_p$. This axis is chosen because the permanent dipole moment of the $(PO_2)^-$ group points along this direction. Electric fields originate from water dipoles, the charged ions, and the $(PO_2)^-$ groups, all being included in the simulations. Electric fields for DNA and RNA were evaluated for all $(PO_2)^-$ groups at 800 equally spaced snapshots during the last 8 ns of the TIP4P-FB/12-6-4 LJ trajectory, electric fields imposed on $(PO_2)^-$ groups of $tRNA^{Phe}$ consider 1600 equally spaced snapshots during the last 16 ns. Average electric fields used for the field-frequency correlation of solvent exposed $(PO_2)^-$ groups were averaged over the last 2 ns interval for RNA while for $tRNA^{Phe}$ electric fields were averaged over the entire 16 ns interval, indicating a higher fluxionality of the hydration shell of RNA compared to folded $tRNA^{Phe}$.

Vibrational frequencies were simulated on the mixed quantum mechanical/molecular mechanical (QM/MM) level of theory as described in detail in ref 29, taking into account the first solvation shell around $(PO_2)^-$ groups and contact ions in the QM region. QM/MM simulations employed the B3LYP-D method (basis: 6-31G* and 6-311+G** for phosphorus atoms) and were performed with the NWchem program.[38]

# 3. RESULTS

**3.1. Linear Infrared Absorption Spectra.** Infrared absorption spectra of the different molecular systems in water are summarized in Figure 1. The absorbance $A$ =

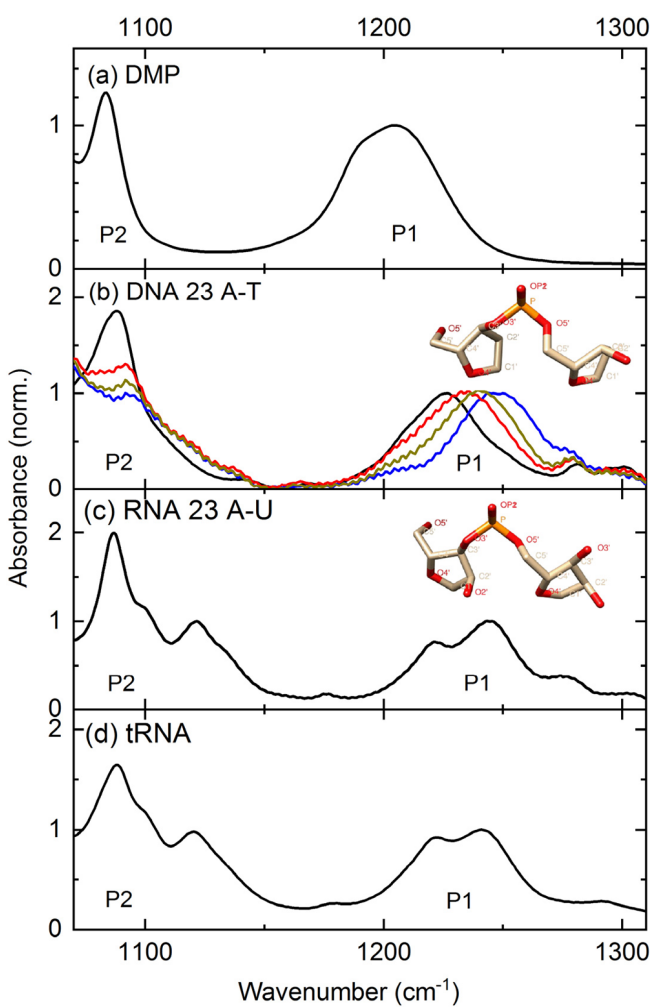

**Figure 1.** Infrared absorption bands P1 (normalized to the peak value) and P2 of the asymmetric and symmetric $(PO_2)^-$ stretching bands of aqueous solutions of (a) dimethyl-phosphate (DMP, concentration $c$ = 200 mM), (b) double-stranded DNA oligomers containing 23 alternating A-T base pairs ($c$ = 10 mM, black line), (c) double-stranded RNA oligomers containing 23 alternating A-U base pairs ($c$ = 10 mM), and (d) E.c. tRNA ($c$ = 4 mM). The P1 band of DMP [panel a] consists of two components from different conformers. The colored lines in panel b represent spectra of DNA films at a relative humidity of 75% (red), 33% (dark yellow), and 0% (blue). The substructure of the RNA P1 bands [panels c and d] reflect different hydration geometries of phosphate groups.

$-\log(T)$ ($T$: sample transmission) is plotted as a function of wavenumber and normalized to the peak value of the band P1 of the asymmetric $(PO_2)^-$ stretching vibration $\nu_{AS}(PO_2)^-$. This band arises between 1170 and 1300 cm$^{-1}$ with a total spectral half width (fwhm) on the order of 50 cm$^{-1}$. In the spectrum of DMP in water (Figure 1a), the P1 band consists of two contributions due to the gt conformer (maximum at ~1190 cm$^{-1}$) and the gg conformer (maximum at 1220

cm$^{-1}$).[39] The symmetric $(PO_2)^-$ stretching band P2 is located around 1083 cm$^{-1}$.

Double-stranded DNA oligomers consisting of 23 alternating A-T base pairs were studied both in aqueous solution and in thin films at a reduced hydration level. The liquid sample contained Na$^+$ counterions whereas the DNA films were prepared with cetyltrimethylammonium chloride (CTMA) counterions. The black line in Figure 1b gives the infrared spectrum of DNA in solution with the P1 band at 1226 cm$^{-1}$ and the P2 band at 1088 cm$^{-1}$. At such high hydration level with more than 100 water molecules per phosphate group, the DNA oligomers exist in a B-helix structure.[1] A reduced hydration level of the $(PO_2)^-$ group gives rise to a pronounced blueshift of the P1 band[23,40,41] as is evident from the spectra of DNA film samples (colored lines in Figure 1b). A gradual decrease of the water content to 75% relative humidity (R.H.) (nine waters per phosphate, red line) and 33% R.H. (three waters per phosphate, dark yellow line) shifts the maximum of the P1 band to 1234 and 1240 cm$^{-1}$, respectively. For 0% R.H., which corresponds to an average concentration of less than one water molecule per $(PO_2)^-$, that is, nearly water-free conditions, the P1 band peaks around 1248 cm$^{-1}$.

Peak positions of the P1 band of DNA and RNA are summarized in Figure 2. In Figure 2a, results from the present DNA film data set and literature[40,41] are plotted as a function of the humidity level (upper abscissa scale). The spectral redshift with increasing humidity is a hallmark of the increasing strength of the electric field the hydrating water molecules exert on the $(PO_2)^-$ unit, as will be discussed later. It should be noted that similar but much smaller frequency shifts occur in the range of the P2 band.

The infrared absorption spectrum of an aqueous solution of double-stranded RNA consisting of 23 alternating A-U base pairs is shown in Figure 1c. This short RNA structure forms an A-type helix geometry in solution.[1] The P1 band displays subcomponents, in particular two strong contributions with maxima at 1220 and 1245 cm$^{-1}$ and a weak component at 1275 cm$^{-1}$. The peak position of the P2 band is at 1087 cm$^{-1}$. The P1 and P2 bands are complemented by the absorption of other backbone vibrations, the linker C–O stretch at 1102 cm$^{-1}$, the C2'–OH stretch at 1120 cm$^{-1}$, and the ribose C1'–O4'–C4' stretch at 1133 cm$^{-1}$.[22]

Figure 1d displays the infrared absorption of E.c. tRNA which contains some 80 nucleotides in a folded cloverleaf structure.[42] In protein synthesis, tRNA decodes the information provided by mRNA via its anticodon units and selects the corresponding amino acid to be added to a protein molecule. The folded tRNA structure exhibits different single-stranded loop sections and double-helical parts. The main features of the infrared absorption spectrum are similar to that of the short RNA duplex (Figure 1c) with somewhat different relative amplitudes of the P1 and P2 bands. In particular, there are three components of the P1 band.

The water shells of DNA and RNA contain solvated counterions, in the present study Na$^+$ or K$^+$ ions, which are sources of an electric field as well. There is variety of solvation geometries, including ions solvated by a neat water shell, solvent-separated ion pairs in which a phosphate group and a counterion are separated by a single water layer, and contact ion pairs with phosphate and counterion in direct proximity.[6–8] In folded RNA structures, for example, tRNA,[42] the interaction with Mg$^{2+}$ ions is regarded to be particularly relevant for balancing the negative phosphate charges and, thus, stabilizing the macromolecular structures electrostatically. To study the impact of Mg$^{2+}$ ions on the $\nu_{AS}(PO_2)^-$ vibrations, we measured infrared spectra of DMP, RNA, and E.c. tRNA in aqueous solutions containing a defined excess concentration of Mg$^{2+}$ ions.[29,43,44] In Figure 3, spectra are shown for different values of the quantity $R = c(Mg^{2+})/c(s)$, the ratio of the Mg$^{2+}$ concentration $c(Mg^{2+})$ relative to that of s = DMP, RNA, or tRNA. In all cases, one observes a decrease of the main absorption components with increasing $R$ and the formation of a blue-shifted absorption band between 1240 and 1270 cm$^{-1}$ for DMP (Figure 3a) and around 1275 cm$^{-1}$ for the two RNA samples (Figure 3b,c). To better illustrate the changes in the RNA spectra, we present differential absorption spectra of E.c. tRNA (Figure 3d) in which the absorption without Mg$^{2+}$ ions ($R = 0$) has been subtracted from the spectra recorded with Mg$^{2+}$ excess ions ($R \neq 0$). The differential spectra clearly exhibit the formation of the blue-shifted band with a strength proportional to R. A similar behavior has been observed with tRNA$^{Phe}$ from yeast.[29]

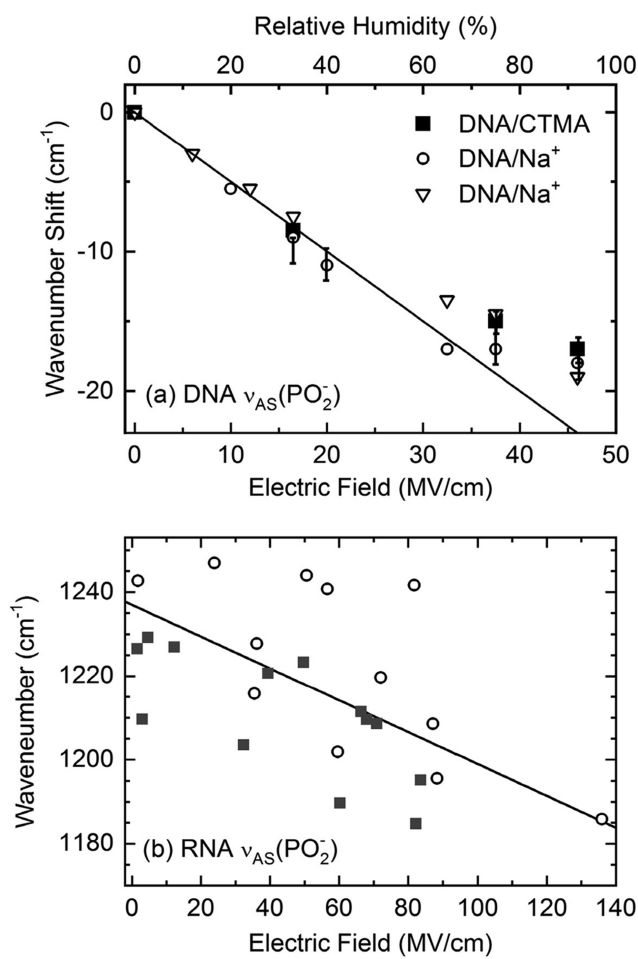

**Figure 2.** Frequency position of the $\nu_{AS}(PO_2)^-$ band as a function of the electric field acting on the $(PO_2)^-$ group. (a) Measured shift $\Delta\nu = \nu_{AS}(R.H.) - \nu_{AS}(R.H.= 0\%)$ of the $\nu_{AS}(PO_2)^-$ frequency position of different DNA samples (symbols) as a function of relative humidity (R.H., upper abscissa scale). Data are taken from refs 23 (squares), 40 (circles), and 41 (triangles). The electric field scale on the bottom abscissa and the solid line were calculated with a frequency tuning rate of $-0.5$ cm$^{-1}$/(MV/cm) assuming a zero electric field at 0% R.H. (b) Frequency positions on QM/MM level of theory of solvent exposed $\nu_{AS}(PO_2)^-$ of tRNA$^{Phe}$ (open symbols) and double-stranded RNA (solid symbols) as a function of the electric field derived from MD simulations. The different electric field values correspond to different local hydration geometries. The solid line gives the result expected for a tuning rate of $-0.38$ cm$^{-1}$/(MV/cm).

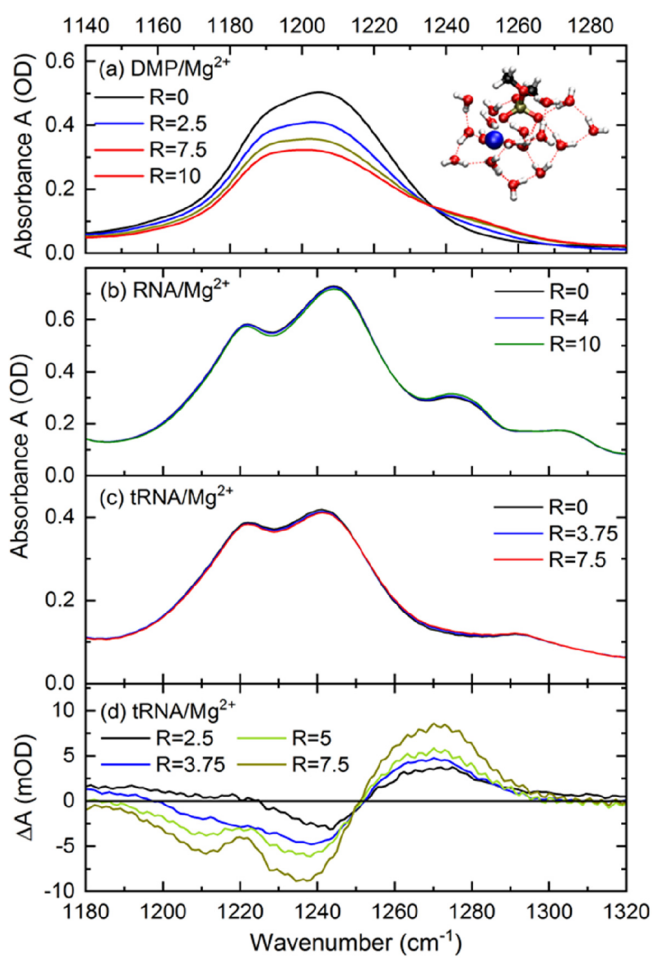

**Figure 3.** Linear infrared absorption spectra of asymmetric phosphate stretching vibrations $\nu_{AS}(PO_2)^-$ of aqueous solutions of (a) DMP (concentration c = 200 mM), (b) double-stranded RNA oligomers containing 23 alternating A-U pairs (c = 6.9 mM), and (c) E.c. tRNA (c = 4 mM), containing additional Mg$^{2+}$ ions. The quantity $R = c(Mg^{2+})/c(s)$ is the ratio of Mg$^{2+}$ concentration to the concentration $c(s)$ of s = DMP, RNA, and tRNA. Spectra are shown for the range from $R = 0$ (without Mg$^{2+}$) to $R = 10$. In all cases, a blue-shifted absorption component is observed upon Mg$^{2+}$ addition. (d) Differential absorbance spectra $\Delta A = A(c(Mg^{2+})) - A_0$ of E.c. tRNA for different Mg$^{2+}$ excess concentrations ($A(c(Mg^{2+}))$, absorbance with Mg$^{2+}$ excess; $A_0$, absorbance without Mg$^{2+}$ ions). The rise of absorption around 1270 cm$^{-1}$ is a hallmark of the formation of Mg$^{2+}$-phosphate contact ion pairs.

The blue-shifted absorption bands are due to $(PO_2)^-$ groups which form contact ion pairs (CIPs) with Mg$^{2+}$ ions.[25,29,43,44] In the prevailing contact geometry, one of the $(PO_2)^-$ oxygen atoms is incorporated in the octahedral first solvation shell of the Mg$^{2+}$ ion. A quantitative analysis of the DMP/Mg$^{2+}$ spectrum for $R = 10$ gives a fraction of DMP molecules forming CIPs on the order of 30%.[44] For RNA at $R = 10$ and tRNA at $R = 7.5$, the number of CIPs per RNA molecule is between 2 and 3.[29]

**3.2. Two-Dimensional Infrared Spectra.** The structure fluctuations of the molecular ensembles in a broad time range lead to fluctuations of the direction and strength of the electric field the environment exerts on the $(PO_2)^-$ groups. The concomitant changes of the vibrational potential energy surfaces give rise to spectral diffusion and affect the vibrational frequencies and line shapes directly. To get insight in such processes, we measured 2D-IR spectra of DMP, DNA, RNA, and tRNA in water with and without additional ionic species.

The 2D-IR spectra of DMP without addition of $Mg^{2+}$ ($R = 0$, Figure 4a) consist of an elliptic peak on the v = 0−1

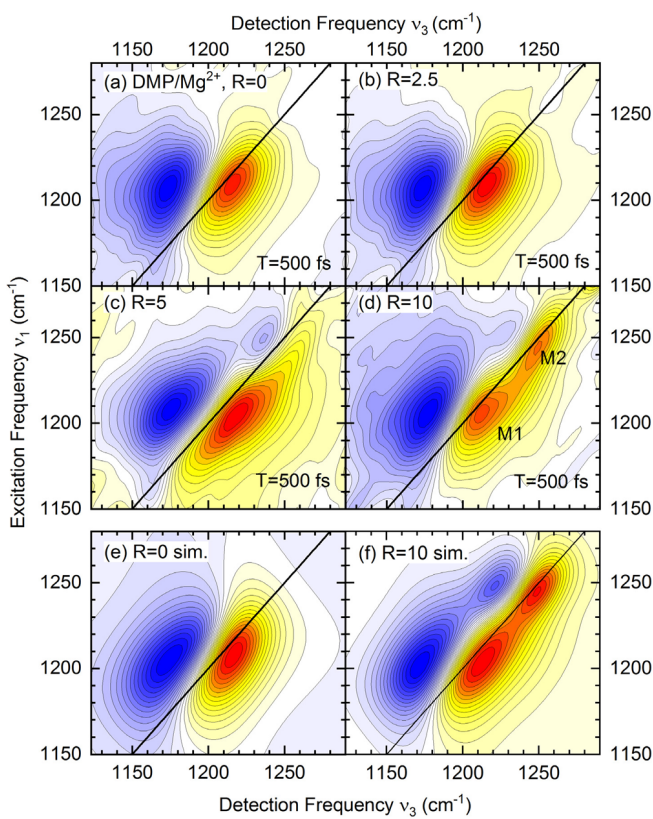

Figure 4. Two-dimensional infrared (2D-IR) spectra of DMP in water in the range of the asymmetric phosphate stretching ($\nu_{AS}(PO_2)^-$) band recorded at a waiting time $T = 500$ fs. The absorptive 2D signal is plotted as a function of the excitation frequency $\nu_1$ and the detection frequency $\nu_3$. Yellow-red contours represent signals due to the fundamental (v = 0−1) transition, blue contours the excited state v = 1−2 absorption. The signal change between neighboring contour lines is 7.5%, the solid black lines are the frequency diagonals $\nu_1 = \nu_3$. The quantity $R = c(Mg^{2+})/c(DMP)$ is the ratio of $Mg^{2+}$ to DMP concentration. (a) The 2D-IR spectrum of DMP in neat water ($R = 0$). (b−d) The 2D-IR spectra for $R = 2.5$, $5$, and $10$. With increasing $R$, a blue-shifted 2D peak arises which is due to DMP/$Mg^{2+}$ contact ion pairs. (e,f) Simulated 2D-IR spectra for $R = 0$ and $R = 10$.

transition of the $\nu_{AS}(PO_2)^-$ mode of the gt and gg conformers (yellow-red contour) and a more extended contribution at smaller detection frequencies $\nu_3$ (blue contour) which is due to the v = 1 to 2 transitions. The redshift of the latter peak along $\nu_3$ reflects the (diagonal) anharmonicity of the $\nu_{AS}(PO_2)^-$ mode. The elliptic shape of the v = 0−1 peak points to a limited inhomogeneous broadening caused by the structural diversity of hydration geometries around the $(PO_2)^-$ groups. The width of the contour along the antidiagonal direction perpendicular to the diagonal ($\nu_1 = \nu_3$) line is dominated by lifetime broadening, reflecting the short 300 fs decay of the v = 1 state of the $\nu_{AS}(PO_2)^-$ vibration.[43,44] With increasing $Mg^{2+}$ content, i.e., increasing R, the 2D-IR spectra develop an additional contribution to the yellow-red contours at higher detection frequency (Figure 4b-d), most clearly manifested in the two separate peaks M1 and M2 observed for $R = 10$ (Figure 4d). There are no cross peaks in the 2D-IR spectra of Figure 4(b-d). This fact shows that the peaks M1 and M2 are due to different uncoupled phosphate environments, one consisting of neat water (M1) and the other representing CIPs with a $Mg^{2+}$ ion. Pump−probe experiments reported in refs 43 and 44 give a vibrational lifetime of the M1 and the M2 component of 300 fs and ∼400 fs, respectively.

The 2D-IR spectra of DMP in water were simulated for $R = 0$ and $R = 10$ with the help of a density matrix approach for describing the third-order nonlinear response of the sample.[45,46] The v = 0−1 and v = 1−2 transitions of the $\nu_{AS}(PO_2)^-$ mode of three different species were included, the gt and gg conformers of DMP and DMP/$Mg^{2+}$ CIPs. A key ingredient for describing the line shapes of the linear infrared absorption and 2D-IR spectra is the frequency fluctuation correlation function (FFCF) which was approximated by a sum of two Kubo terms, the first decaying with a 300 fs time constant and the second slower one with a 50 ps time constant. In the analysis of the spectra, such time constants were fixed and the respective frequency fluctuation amplitudes $\Delta\nu_1$ and $\Delta\nu_2$ were varied to fit the experimental spectra. Details of this approach have been given in ref 21. The calculated 2D-IR spectra presented in Figure 4e for $R = 0$ and Figure 4f for $R = 10$ are in good agreement with their experimental counterparts in Figure 4a,d. The parameters of the calculation are listed in Table 1.

Figure 5 summarizes selected 2D-IR spectra of DNA, RNA, and tRNA. The spectrum of DNA in neat water (Figure 5a) exhibits a single peak on the v = 0−1 transition centered around 1230 $cm^{-1}$ (yellow-red contour) in line with the linear infrared absorption spectrum (Figure 1b, black line). The elliptic shape of the 2D peak is due to a pronounced inhomogeneous broadening which reflects variations in the local hydration geometry and, thus, electric field acting on the $(PO_2)^-$ groups. The peak red-shifted along $\nu_3$ (blue contour) is again due to the v = 1−2 transition. The spectrum of RNA in neat water (Figure 5b) shows a v = 0−1 signal (yellow-red) which consists of three different components. The latter are

**Table 1. Parameter Values from the Numerical Simulation of the 2D-IR Spectra of DMP in Water (Figure 4e,f)$^a$**

| $\nu_{AS}(PO_2)^-$ | lifetime (v = 1) state (fs) | relative amplitude $R = 0/10$ | frequency $\nu_i$ ($cm^{-1}$) | $\Delta_{Diagonal}$ ($cm^{-1}$) $R = 0/10$ | $\Delta\nu_{1i}$ ($cm^{-1}$) $\tau = 300$ fs $R = 0/10$ | $\Delta\nu_{2i}$ ($cm^{-1}$) $\tau = 50$ ps $R = 0/10$ |
|---|---|---|---|---|---|---|
| DMP (gt) | 300 | 90/83 | 1195 | 26/26 | 17/15 | 9.5/8 |
| DMP gg | 300 | 100/100 | 1215 | 21/25 | 13.8/13.8 | 10.7/11.7 |
| DMP/$Mg^{2+}$ | 500 | 0/49 | 1248.5 | −/25 | −/10.6 | −/10.0 |

$^a$The fourth column gives the v = 0−1 transition frequency of the respective $\nu_{AS}(PO_2)^-$ vibration and the fifth column its diagonal anharmonicity $\Delta_{Diagonal}$. The quantities $\Delta\nu_1$ and $\Delta\nu_2$ are the frequency fluctuation amplitudes of the fast and slow Kubo term in the frequency fluctuation correlation function. The quantity $R = c(Mg^{2+})/c(DMP)$ represents the ratio of $Mg^{2+}$ to DMP concentration.

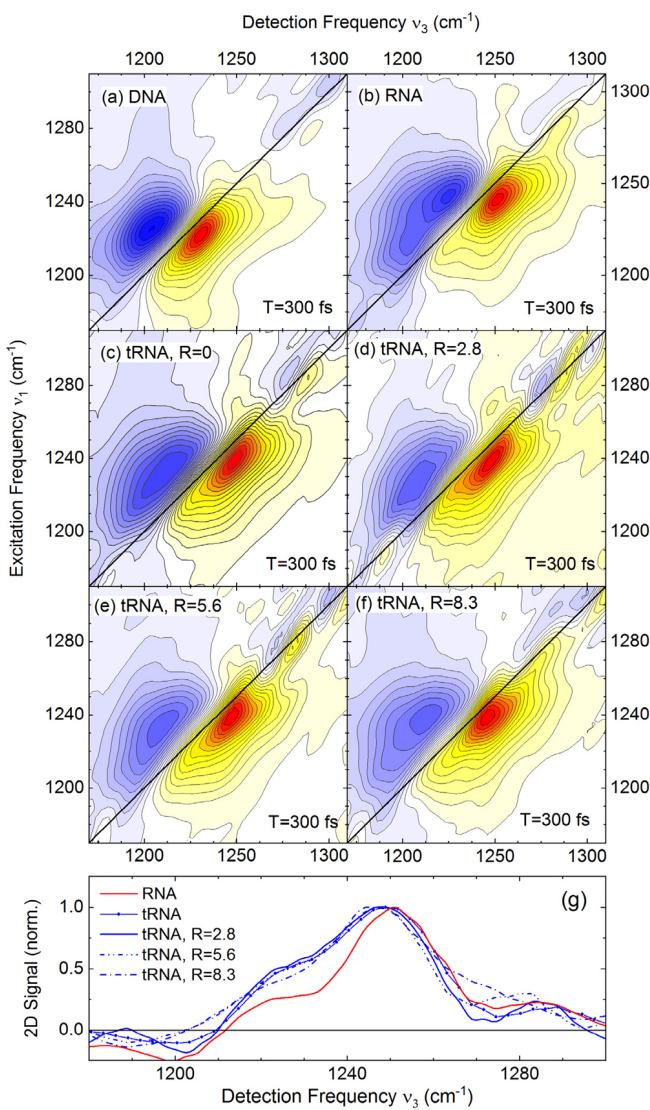

**Figure 5.** Two-dimensional infrared (2D-IR) spectra of $\nu_{AS}(PO_2)^-$ vibrations of (a) double-stranded DNA oligomers with alternating A-T base pairs and (b) double-stranded RNA oligomers with alternating A-U base pairs in water. The absorptive 2D signal is plotted as a function of the excitation frequency $\nu_1$ and the detection frequency $\nu_3$. Yellow-red contours represent signals on the fundamental (v = 0−1) transition, blue contours on the v = 1−2 transition. The signal change between neighboring contour lines is 7.5%. The black solid line is the frequency diagonal $\nu_1 = \nu_3$. (c−f) The 2D-IR spectra of E.c. tRNA in water for different excess concentrations of $Mg^{2+}$ ions. The quantity $R = c(Mg^{2+})/c(tRNA)$ is the ratio of $Mg^{2+}$ to tRNA concentration. (g) Cuts of the 2D-IR spectra of RNA and tRNA along a diagonal line through $(\nu_1,\nu_3) = (1242, 1250)\ cm^{-1}$ (RNA) and $(1240, 1250)\ cm^{-1}$ (E.c. tRNA).

clearly visible in the diagonal cut of this peak parallel to the $\nu_1 = \nu_3$ line (Figure 5g). The strongest component centered around 1250 cm$^{-1}$ is complemented by two weaker components at 1220 and 1280 cm$^{-1}$ in analogy to the infrared absorption spectrum in Figure 1c. There are no cross peaks between the different contributions which correspond to different hydration geometries of RNA phosphate groups. The absence of cross peaks is particularly evident from cuts of the spectra along detection frequency $\nu_3$ which have been presented in ref 25. A very similar behavior is observed for

E.c. tRNA in neat water, both in the 2D-IR spectrum (Figure 5c) and the corresponding diagonal cut (Figure 5g).

The addition of $Mg^{2+}$ ions to the E.c. tRNA sample leads to an enhancement of the 2D-IR signal strength between 1265 and 1280 cm$^{-1}$ (Figure 5d−f), reflecting the formation of CIPs. While the differential linear absorption spectra in Figure 1d display a similar behavior, the relative strength of the CIP component is substantially higher in the 2D-IR spectra. This enhancement is mainly due to the longer vibrational lifetime of this excitation which is on the order of 800 fs, compared to the 300 fs lifetimes of the two other components. At a population time of $T = 300$ fs at which the spectra in Figure 5 were recorded, the shorter-lived signal contributions thus show a reduced relative amplitude, facilitating the observation of the blue-shifted band.

**3.3. Theory Results.** The spectral shift of the P1 and P2 bands upon applying an electric field are described in an approximate way by the so-called Stark tuning rate, giving the frequency shift of a vibrational band per unit electric field $E_p$. Stark tuning rates for the $(PO_2)^-$ stretching vibrations have been reported in refs 26 and 39 with the former benchmarked by Stark shift experiments with static electric fields. Such work gives tuning rates between −0.4 and −0.54 cm$^{-1}$/(MV/cm). Applying a tuning rate of −0.5 cm$^{-1}$/(MV/cm) to the spectral peak positions of the P1 band of DNA, we derive the electric field scale for $E_p$ plotted on the bottom abscissa of Figure 2a and the straight solid line for the electric-field dependent vibrational frequency. We recall that the $E_p$ values represent the projection of the total electric field $E$ from the hydrating water molecules on the bisector of the O1═P═O2 angle, the direction of the permanent $(PO_2)^-$ dipole moment. The zero point of this scale is chosen for the frequency position of the P1 band of DNA at 0% R.H., corresponding to less than one water molecule per phosphate group. With this assumption, one derives a maximum electric field on the order of 50 MV/cm.

In Figure 2b, we analyze the correlation of electric field values derived from MD simulations with QM/MM derived vibrational frequencies $\nu_{AS}(PO_2)^-$ at solvent exposed surface positions of tRNA$^{Phe}$ (open symbols) and double-stranded RNA (solid symbols) as a function of the temporally averaged projected electric fields $E_p$. Electric fields cover an $E_p \sim 100$ MV/cm range and $\nu_{AS}(PO_2)^-$ vibrational frequencies cover an ~65 cm$^{-1}$ range, in good agreement with the experimental width (Figure 1c,d). Compared to DNA, the $\nu_{AS}(PO_2)^-$ frequency positions at a particular value of $E_p$ are scattered over a larger range, a consequence of the more heterogeneous hydration structure of RNA. Nevertheless, a field-frequency correlation employing a tuning rate of −0.38 cm$^{-1}$/(MV/cm) (Figure 2b, best linear fit) provides a reasonable correlation for the entire range of $\nu_{AS}(PO_2)^-$ vibrational frequencies.

Figure 6 provides an analysis of electric field distributions and hydration geometries for $(PO_2)^-$ groups of fully hydrated double-stranded DNA, double-stranded RNA, and tRNA$^{Phe}$. For a prototypical $(PO_2)^-$ group of DNA we find that the total imposed electric field is on average $E \sim 70$ MV/cm (Figure 6a). The absolute magnitude is reduced to $E_p \sim 50$ MV/cm upon projection on the O1═P═O2 angle bisector axis which reflects the approximate directional alignment of the electric field vector imposed by $H_2O$ molecules and the O1═P═O2 angle bisector axis. The standard deviation of the field distribution is 20.5 MV/cm. An instantaneous hydration geometry around the $(PO_2)^-$ group of DNA is shown in

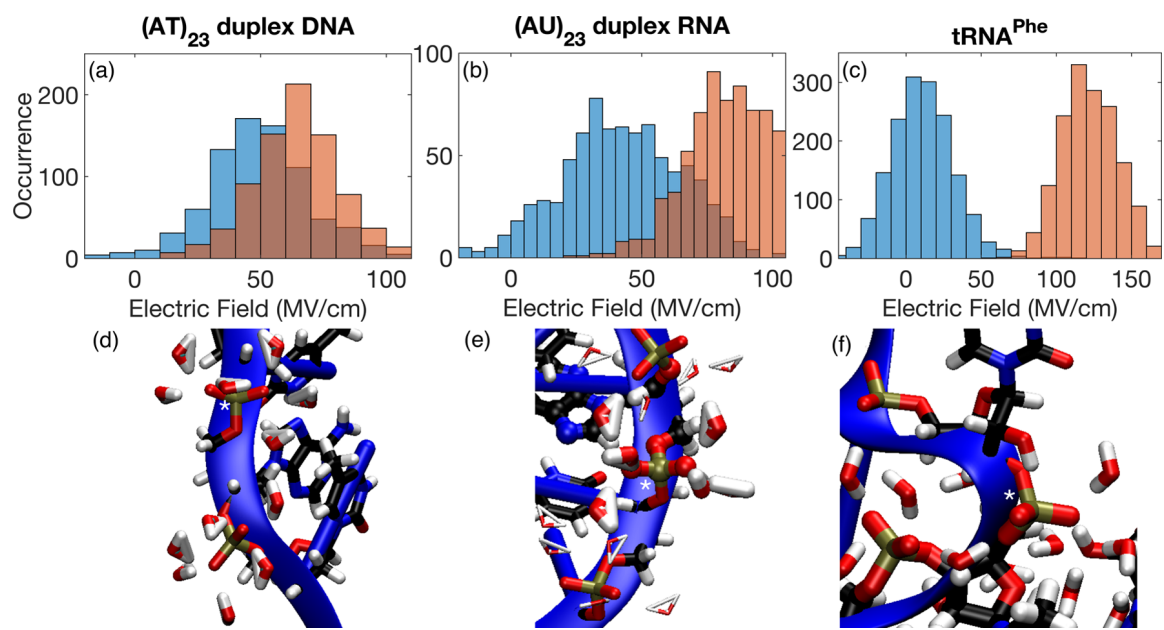

**Figure 6.** Electric field distributions and molecular hydration structures of (a,d) $(AT)_{23}$ duplex DNA, (b,e) $(AU)_{23}$ duplex RNA, and (c,f) tRNA$^{Phe}$ from molecular dynamics simulations. The electric field distributions in panels (a−c) were evaluated at the $(PO_2)^-$ group indicated with an asterisk in panels (d−f) and are given as the magnitude of the total electric field at the bisector midpoint of the $(PO_2)^-$ group (red bars) followed by projection on the angle bisector axis of the O1=P=O2 unit (blue bars). The electric field values are taken from snapshots during the last part of the respective MD trajectory. The molecular hydration structures show instantaneous snapshots of $H_2O$ molecules within 2.7 Å of phosphate O1 and O2 atoms forming the first hydration shell around the $(PO_2)^-$ group.

Figure 6d. We identify a first hydration shell of six $H_2O$ molecules with the typical tetrahedral coordination of O1 and O2 atoms with the $(PO_2)^-$ group. Because of the ~6 Å separation of neighboring $(PO_2)^-$ groups, hydration shells are independent and do not overlap significantly.[5]

For double-stranded RNA and tRNA$^{Phe}$ (Figure 6b,c), the magnitude of the total electric field imposed on the $(PO_2)^-$ group is substantial ($E \sim 80-150$ MV/cm) and can even exceed the magnitude found for DNA. Nevertheless, upon projection on the O1=P=O2 angle bisector axis $E_p$ is greatly reduced, for example, to a mean field of $E_p \sim 11.2$ MV/cm for tRNA$^{Phe}$. Thus, the direction of the total electric field imposed on the $(PO_2)^-$ group is, due to the steric constraints of the RNA surface, particularly different from the DNA case and leads to a substantial reduction upon projection. The origin of the asymmetric field direction imposed on the $(PO_2)^-$ group can be rationalized with the help of the instantaneous hydration structures shown in Figure 6e,f. For double-stranded RNA (Figure 6e), the symmetry between O1 and O2 atoms of the $(PO_2)^-$ group is broken. The O1 oxygen atom points toward the solvent and is surrounded by three water molecules in a tetrahedral arrangement. In contrast, the O2 oxygen atom points toward the deep, spatially constrained minor groove. Such steric constraints and the short ~4.6 Å separation of neighboring $(PO_2)^-$ groups facilitate the formation of $H_2O$ bridges between $(PO_2)^-$ groups. Thus, the hydration shells between neighboring $(PO_2)^-$ groups overlap due to an economization of the number of hydrogen bonds with a reduced number of $H_2O$ molecules. Moreover, the number of $H_2O$ molecules of the second hydration shell is reduced within the groove, compared to the freely accessible solvent volume around the O1 atoms.

For tRNA$^{Phe}$, with the spatial constraints determined by the folded tertiary structure only four $H_2O$ molecules are part of the first solvation shell of the $(PO_2)^-$ group that is completed by hydroxyl groups of the sugar−phosphate backbone. Similar to the case of double-stranded RNA, the reduced number of $H_2O$ molecules and short inter-$(PO_2)^-$ distances lead to the formation of $H_2O$ bridges between $(PO_2)^-$ groups. The asymmetric arrangement of $H_2O$ molecules in the first solvation shell of the O1 and O2 atoms of the $(PO_2)^-$ groups lowers the electric field $E_p$ projected on the O1=P=O2 angle bisector axis.

## 4. DISCUSSION

The experimental results demonstrate surprisingly similar spectroscopic and dynamic properties of $\nu_{AS}(PO_2)^-$ excitations in markedly different molecular systems, including DMP, short DNA, and RNA duplexes, much longer DNA double helices (discussed in refs 24,40, and 41) and tRNA structures. In all cases, one finds a vibrational lifetime of $\nu_{AS}(PO_2)^-$ vibrations of approximately 300 fs, pointing to a predominant relaxation via anharmonically coupled vibrations of the phosphate group at lower frequency.[23] Upon addition of $Mg^{2+}$ ions, the formation of contact ion pairs leads to a pronounced blueshift of the $\nu_{AS}(PO_2)^-$ fundamental transition, accompanied by an increase of the vibrational lifetime to 600−800 fs compared to the 300 fs lifetime. The increase of lifetime may reflect an increased detuning of the v = 1 state of the $\nu_{AS}(PO_2)^-$ mode from the accepting vibrational energy level in the relaxation process.

We now discuss the mechanisms inducing the pronounced frequency shifts of the $\nu_{AS}(PO_2)^-$ vibrations. Compared to the $\nu_{AS}(PO_2)^-$ transition frequency in a water-free environment, addition of a hydration shell with embedded positively charged ions causes a frequency red-shift. In contrast, CIPs of $(PO_2)^-$ groups and $Mg^{2+}$ ions embedded in water display a blue-shifted $\nu_{AS}(PO_2)^-$ frequency. The experimental data and theoretical results for the $\nu_{AS}(PO_2)^-$ mode in the absence of CIPs suggest a common interaction mechanism behind the observed red-shifts. Underlying this mechanism is a modification of the electronic structure of the highly polarizable $(PO_2)^-$ group by the external electric field from the dipole moments of water molecules and, to a much lesser extent, solvated positive ions. The change of electronic structure leads to a softening of the vibrational potentials of the symmetric and asymmetric $(PO_2)^-$ stretching vibrations and, thus, a redshift of their fundamental $v = 0-1$ transition with increasing electric field. Our calculations show that the frequency shifts scale roughly linearly with $E_p$, the projection of the total external electric field E on the direction of the permanent $(PO_2)^-$ dipole, the bisector of the O1=P=O2 axis. This behavior suggests a predominance of the dipole-mediated interaction with the solvent field over higher-order multipole terms in the total electrostatic energy. It should be noted that the hydrogen bonds between $(PO_2)^-$ and water molecules are dominated by electrostatic interactions contributing to $E_p$, while the role of covalent and charge transfer interactions is negligible.[39]

The relevant range of time-averaged electric fields $E_p$ is between ∼10 and 100 MV/cm and the maximum redshift on the order of $\Delta\nu = 65$ cm$^{-1}$. For DNA (Figure 2a), there is a linear decrease of vibrational frequency with an increasing hydration level up to a relative humidity (R.H.) around 60%. In this range, the successive addition of water molecules results in a linear increase of the electric field $E_p$ acting on the $(PO_2)^-$ group which is well represented by a Stark tuning rate of −0.5 cm$^{-1}$/(MV/cm) (solid line in Figure 2a). This tuning rate is in line with theoretical calculations.[26,39] The observation of a 1248 cm$^{-1}$ band in dehydrated DNA with less than one water molecule per phosphate group, that is, a minor concentration of water dipoles and, thus, negligible $E_p$ supports this picture. A relative humidity of 60% corresponds to a number of six water molecules per nucleotide,[47] identical to the number of $H_2O$ molecules in the first solvation shell around a $(PO_2)^-$ group. In agreement with the theoretical analysis, this observation suggests that water molecules in the tetrahedral first solvation shell represent the main source of $E_p$ which is additive in the number of water molecules. Further addition of water has less impact on the $\nu_{AS}(PO_2)^-$ frequency, as is evident from the frequency positions observed at 70 and 92% R.H. The latter clearly deviate from the linear behavior, in line with more detailed data reported in ref 40.

The correlation between the $\nu_{AS}(PO_2)^-$ frequencies of RNA and the local electric field $E_p$ shows a stronger variation (Figure 2b). On the one hand, the larger dispersion can potentially arise from methodological limitations of the current approach where the electric field is evaluated on an electrostatic level that facilitates extensive sampling of field distributions (Figure 6) while vibrational frequencies rely on the numerically costly QM/MM level of theory that accounts for higher order electrostatic contributions as well as covalent and charge transfer interactions.[39] On the other hand, particularities of the larger variety of hydration geometries and microscopic details of the hydration shell in double-stranded RNA are not fully accounted for in the current electric field-dependent mapping of vibrational frequencies. For example, the 2′-OH group of the ribose units can impose ordered hydration structures[22] with a strong impact on the water structure. Our QM/MM simulations indicate that such structures can affect the mode mixing between the 2′-OH and the $(PO_2)^-$ group. Such microscopic complexity goes beyond an electric field-depend-ent mapping of the vibrational frequencies and will be addressed in detail in future work.

The ∼1245 cm$^{-1}$ band in double-stranded RNA and tRNA$^{Phe}$ is due to $(PO_2)^-$ groups in a highly asymmetric hydration shell generating a small field $E_p$ only. $H_2O$ molecules forming bridges between neighboring, closely spaced $(PO_2)^-$ groups, for example, of an A-helix, contribute to the asymmetry of the hydration shell but are not instrumental for the observed ∼1245 cm$^{-1}$ band. In simulations of the asymmetric $(PO_2)^-$ stretching vibrations of double-stranded RNA and tRNA$^{Phe}$, we find strong local frequency variations with neighboring $(PO_2)^-$ groups bridged by $H_2O$ molecules contributing at ∼1220 and ∼1245 cm$^{-1}$.

On the other hand, the $(PO_2)^-$ groups with a tetrahedral hydration shell with the up to six water molecules largely symmetrically arranged around both oxygen atoms of the $(PO_2)^-$ group exist in double stranded RNA and tRNA, experience a high $E_p$, and give rise to $\nu_{AS}(PO_2)^-$ absorption around 1220 cm$^{-1}$, similar to DNA. In the infrared spectra of the fully hydrated RNA systems, the parallel occurrence of the two hydration patterns gives rise to the bimodal shape with peaks at 1220 and 1245 cm$^{-1}$.

All infrared spectra in Figure 1 display distinct absorption bands of the symmetric and asymmetric $(PO_2)^-$ stretching vibrations with a spectral separation of 130−160 cm$^{-1}$. Such normal modes of the $(PO_2)^-$ group represent linear combinations of the two local P=O stretching modes which are coupled via the intramolecular vibrational potential. The frequency separation of the two $(PO_2)^-$ normal modes roughly corresponds to twice the intramolecular coupling which is on the order of 65−80 cm$^{-1}$. This coupling exceeds the field-induced shift of the $\nu_{AS}(PO_2)^-$ vibrations and the coupling of the local oscillators to the external electric field. As a result, there is no field-induced decoupling of the two local oscillators and the pattern of symmetric and asymmetric $(PO_2)^-$ absorption lines is preserved in the infrared spectra.

The electric field from the water environments displays fluctuations on a time scale from 50 fs to several picoseconds due to thermally activated low-frequency motions of water molecules. The numerical analysis of the 2D-IR spectra of DMP allows for estimating fluctuation amplitudes on a subpicosecond time scale with the help of Stark tuning rates. The frequency fluctuation amplitudes $\Delta\nu_1$ of 10−18 cm$^{-1}$ in the FFCF translate into fluctuation amplitudes $\Delta E_p = 20−36$ MV/cm when applying the frequency tuning rate of 0.5 cm$^{-1}$/(MV/cm).[39] Similar values of $\Delta\nu_1$ and, thus, $\Delta E_p$ have been found from an analysis of the 2D-IR spectra of DNA[21] and tRNA$^{Phe}$.[29] An analysis of fast fluctuations in the MD trajectories gives very similar amplitudes.[39] It should be noted that the electric field distributions shown in Figure 6 are averaged over a much longer time interval of approximately 8−16 ns and correspond to the electric field distribution at a particular site rather than fast fluctuation amplitudes.

Our results provide evidence for the formation of $(PO_2)^-/Mg^{2+}$ CIPs in the different molecular systems. The CIP geometries are characterized by the integration of one of the $(PO_2)^-$ oxygen atoms in the octahedral first water shell around $Mg^{2+}$.[25,43] Here, the O−Mg$^{2+}$ distance of ∼2.1 Å is substantially shorter than the length of a $(PO_2)^-$−water hydrogen bond of approximately 2.7 Å. On top of the electrostatic coupling, repulsive exchange interactions come into play at such short interatomic distances and eventually prevail. As a result, the $\nu_{AS}(PO_2)^-$ excitation probes the repulsive part of the interaction potential as manifested in an asymmetric distortion of the interaction potential and concomitant blue-shift of the vibrational transition. It is important to note that the CIP geometry is more rigid than a neat water shell hydrating the $(PO_2)^-$ group. This property results in fluctuation amplitudes $\Delta \nu_1$ of the CIP component in the 2D-IR spectra which are smaller than for water hydration shells (cf. Table 1 and ref 29). The CIP geometries found in the MD simulations exist for periods longer than 1 $\mu$s while the exchange of water molecules around phosphate groups occurs in a time range around 50 ps. It is important to note that CIP geometries play a central role in stabilizing the tertiary structure of tRNA by reducing the repulsive interactions between phosphate groups in the tRNA backbone.[29]

## 5. CONCLUSIONS

The asymmetric $(PO_2)^-$ stretching vibration $\nu_{AS}(PO_2)^-$ represents a sensitive probe of local electric fields originating from a fluctuating water shell. Its pronounced frequency redshift with electric field strength is roughly linear for electric fields between 0 and some 60 MV/cm, resulting in a Stark tuning rate on the order of $-0.5$ cm$^{-1}$/(MV/cm) in a large class of molecular systems, including DMP, DNA, and different types of RNA. The combination of 2D-IR spectroscopy with MD simulations gives quantitative insight in ultrafast electric field fluctuations with amplitudes between 20 and 35 MV/cm. The observation window for ultrafast structural and the resulting electric field fluctuations is limited by the 300 fs lifetime of the vibration. Beyond probing electrostatic interactions, the $\nu_{AS}(PO_2)^-$ transition frequency is sensitive to repulsive, that is, exchange interactions which come into play in contact pairs of ions with the $(PO_2)^-$ group. Repulsive interactions induce an upshift of the $\nu_{AS}(PO_2)^-$ frequency by up to 30 cm$^{-1}$, generating a distinct vibrational band that can be isolated in 2D-IR spectra and be applied as a quantitative probe of the concentration of contact pairs.

From an experimental point of view, the $\nu_{AS}(PO_2)^-$ vibration and to lesser extent the symmetric $(PO_2)^-$ stretching vibration hold potential for measuring dynamic vibrational Stark shifts induced, for example, by a strong external electric field at the terahertz frequencies of water fluctuations. In theory, a balanced description of electrostatic, polarization, and dispersion interactions and vibrational frequencies from QM/MM calculations are highly relevant for describing structure, dynamics, and electrical properties of hydrated biomolecules.

## ■ AUTHOR INFORMATION

**Corresponding Author**

**Thomas Elsaesser** − *Max-Born-Institut für Nichtlineare Optik und Kurzzeitspektroskopie, Berlin 12489, Germany;* ⊙ orcid.org/0000-0003-3056-6665; Email: elsasser@mbi-berlin.de

**Authors**

**Jakob Schauss** − *Max-Born-Institut für Nichtlineare Optik und Kurzzeitspektroskopie, Berlin 12489, Germany*

**Achintya Kundu** − *Max-Born-Institut für Nichtlineare Optik und Kurzzeitspektroskopie, Berlin 12489, Germany;* ⊙ orcid.org/0000-0002-6252-1763

**Benjamin P. Fingerhut** − *Max-Born-Institut für Nichtlineare Optik und Kurzzeitspektroskopie, Berlin 12489, Germany;* ⊙ orcid.org/0000-0002-8532-6899

Complete contact information is available at:

https://pubs.acs.org/10.1021/acs.jpcb.1c01502

**Notes**
The authors declare no competing financial interest.

## ■ ACKNOWLEDGMENTS

We thank our former co-workers Ł. Szyc, M. Yang, R. Costard, T. Siebert, and E. M. Bruening for their contributions to the results discussed here and Janett Feickert for expert technical support. This research has received funding from the European Research Council (ERC) under the European Union's Horizon 2020 research and innovation program (Grant Agreements 833365 and 802817).

## ■ REFERENCES

(1) Saenger, W. *Principles of Nucleic Acid Structure*; Springer: Berlin, 1984; Chapter 17.

(2) Laage, D.; Elsaesser, T.; Hynes, J. T. Water Dynamics in the Hydration Shells of Biomolecules. *Chem. Rev.* **2017**, *117*, 10694−10725.

(3) Vlieghe, D.; Turkenburg, J. P.; van Meervelt, L. B-DNA at Atomic Resolution Reveals Extended Hydration Patterns. *Acta Crystallogr., Sect. D: Biol. Crystallogr.* **1999**, *55*, 1495−1502.

(4) Egli, M.; Portmann, S.; Usman, N. RNA Hydration: A Detailed Look. *Biochemistry* **1996**, *35*, 8489−8494.

(5) Schneider, B.; Patel, K.; Berman, H. M. Hydration of the Phosphate Group in Double-Helical DNA. *Biophys. J.* **1998**, *75*, 2422−2434.

(6) Draper, D. E.; Grilley, D.; Soto, A. M. Ions and RNA Folding. *Annu. Rev. Biophys. Biomol. Struct.* **2005**, *34*, 221−243.

(7) Lipfert, J.; Doniach, S.; Das, R.; Herschlag, D. Understanding Nucleic Acid-Ion Interactions. *Annu. Rev. Biochem.* **2014**, *83*, 813−841.

(8) Auffinger, P.; Westhof, E. Water and Ion Binding around RNA and DNA (C,G) Oligomers. *J. Mol. Biol.* **2000**, *300*, 1113−1131.

(9) Anderson, C. F.; Record, M. T. Polyelectrolyte Theories and their Applications to DNA. *Annu. Rev. Phys. Chem.* **1982**, *33*, 191−222.

(10) Fogolari, F.; Brigo, A.; Molinari, H. The Poisson-Boltzmann Equation for Biomolecular Electrostatics: A Tool for Structural Biology. *J. Mol. Recognit.* **2002**, *15*, 377−392.

(11) Gebala, M.; Herschlag, D. Quantitative Studies of an RNA Duplex Electrostatics by Ion Counting. *Biophys. J.* **2019**, *117*, 1116−1124.

(12) Young, M. A.; Jayaram, B.; Beveridge, D. L. Local Dielectric Environment of B-DNA in Solution: Results from a 14 ns Molecular Dynamics Trajectory. *J. Phys. Chem. B* **1998**, *102*, 7666−7669.

(13) Feig, M.; Pettitt, B. M. Sodium and Chlorine Ions as Part of the DNA Solvation Shell. *Biophys. J.* **1999**, *77*, 1769−1781.

(14) Lavery, R.; Maddocks, J. H.; Pasi, M.; Zakrzewska, K. Analyzing Ion Distributions around DNA: Sequence-Dependence of Potassium Ion Distributions from Microsecond Molecular Dynamics. *Nucleic Acids Res.* **2014**, *42*, 8138−8149.

(15) Pan, F.; Roland, C.; Sagui, C. Ion Distributions around Left- and Right-Handed DNA and RNA Duplexes: A Comparative Study. *Nucleic Acids Res.* **2014**, *42*, 13981−13996.

(16) Meisburger, S. P.; Pabit, S. A.; Pollack, L. Determining the Locations of Ions and Water around DNA from X-Ray Scattering Measurements. *Biophys. J.* **2015**, *108*, 2886−2895.

(17) Ermilova, E.; Bier, F. F.; Hölzel, R. Dielectric Measurements of Aqueous DNA Solutions up to 110 GHz. *Phys. Chem. Chem. Phys.* **2014**, *16*, 11256−11264.

(18) Cohen, B. E.; McAnaney, T. B.; Park, E. S.; Jan, Y. N.; Boxer, S. G.; Jan, L. Y. Probing Protein Electrostatics with a Synthetic Fluorescent Amino Acid. *Science* **2002**, *296*, 1700−1703.

(19) Bublitz, G. U.; Boxer, S. G. Stark Spectroscopy: Applications in Chemistry, Biology, and Materials Science. *Annu. Rev. Phys. Chem.* **1997**, *48*, 213−242.

(20) Hithell, G.; Shaw, D. J.; Donaldson, P. M.; Greetham, G. M.; Towrie, M.; Burley, G. A.; Parker, A. W.; Hunt, N. T. Long-Range Vibrational Dynamics are Directed by Watson-Crick Base Pairing in Duplex DNA. *J. Phys. Chem. B* **2016**, *120*, 4009−4018.

(21) Siebert, T.; Guchhait, B.; Liu, Y.; Costard, R.; Elsaesser, T. Anharmonic Backbone Vibrations in Ultrafast Processes at the DNA-Water Interface. *J. Phys. Chem. B* **2015**, *119*, 9670−9677.

(22) Bruening, E. M.; Schauss, J.; Siebert, T.; Fingerhut, B. P.; Elsaesser, T. Vibrational Dynamics and Couplings of the Hydrated RNA Backbone: A Two-Dimensional Infrared Study. *J. Phys. Chem. Lett.* **2018**, *9*, 583−587.

(23) Szyc, Ł.; Yang, M.; Elsaesser, T. Ultrafast Energy Exchange via Water-Phosphate Interactions in Hydrated DNA. *J. Phys. Chem. B* **2010**, *114*, 7951−7957.

(24) Siebert, T.; Guchhait, B.; Liu, Y.; Fingerhut, B. P.; Elsaesser, T. Range, Magnitude, and Ultrafast Dynamics of Electric Fields at the Hydrated DNA Surface. *J. Phys. Chem. Lett.* **2016**, *7*, 3131−3136.

(25) Kundu, A.; Schauss, J.; Fingerhut, B. P.; Elsaesser, T. Change of Hydration Patterns upon RNA Melting Probed by Excitations of Phosphate Backbone Vibrations. *J. Phys. Chem. B* **2020**, *124*, 2132−2138.

(26) Levinson, N. M.; Bolte, E. E.; Miller, C. S.; Corcelli, S. A.; Boxer, S. G. Phosphate Vibrations Probe Local Electric Fields and Hydration in Biomolecules. *J. Am. Chem. Soc.* **2011**, *133*, 13236−13239.

(27) Römer, R.; Hach, R. tRNA Conformation and Magnesium Binding. *Eur. J. Biochem.* **1975**, *55*, 271−284.

(28) Grilley, D.; Soto, A. M.; Draper, D. Direct Quantitation of Mg²⁺- RNA Interactions by Use of a Fluorescent Dye. *Methods Enzymol.* **2009**, *455*, 71−94.

(29) Schauss, J.; Kundu, A.; Fingerhut, B. P.; Elsaesser, T. Magnesium Contact Ions Stabilize the Tertiary Structure of Transfer RNA: Electrostatics Mapped by Two-Dimensional Infrared Spectra and Theoretical Simulations. *J. Phys. Chem. B* **2021**, *125*, 740−747.

(30) Case, D. A.; Ben-Shalom, I.; Brozell, S. R.; Cerutti, D. S.; Cheatham, T.; Cruzeiro, V. W. D.; Darden, T.; Duke, R.; Ghoreishi, D.; Gilson, M. et al. *AMBER 18*; University of California: San Francisco, 2018.

(31) Ivani, I.; Dans, P. D.; Noy, A.; Pérez, A.; Faustino, I.; Hospital, A.; Walther, J.; Andrio, P.; Goñi, R.; Balaceanu, A.; et al. PARMBSC1: A Refined Force-Field for DNA Simulations. *Nat. Methods* **2016**, *13*, 55−58.

(32) Pérez, A.; Marchán, I.; Svozil, D.; Sponer, J.; Cheatham, T. E.; Laughton, C. A.; Orozco, M. Refinement of the AMBER Force Field for Nucleic Acids: Improving the Description of α/γ Conformers. *Biophys. J.* **2007**, *92*, 3817−3829.

(33) Wang, L.-P.; Martinez, T. J.; Pande, V. S. Building Force Fields: an Automatic, Systematic, and Reproducible Approach. *J. Phys. Chem. Lett.* **2014**, *5*, 1885−1891.

(34) Li, P.; Merz, K. M. Taking into Account the Ion-Induced Dipole Interaction in the Nonbonded Model of Ions. *J. Chem. Theory Comput.* **2014**, *10*, 289−297.

(35) Li, P.; Song, L. F.; Merz, K. M. Systematic Parameterization of Monovalent Ions Employing the Nonbonded Model. *J. Chem. Theory Comput.* **2015**, *11*, 1645−1657.

(36) Salomon-Ferrer, R. A.; Götz, W.; Poole, D.; Le Grand, S.; Walker, R. C. Routine Microsecond Molecular Dynamics Simulations with AMBER on GPUs. 2. Explicit Solvent Particle Mesh Ewald. *J. Chem. Theory Comput.* **2013**, *9*, 3878−3888.

(37) Le Grand, S.; Götz, A. W.; Walker, R. C. SPFP: Speed without Compromise—a Mixed Precision Model for GPU Accelerated Molecular Dynamics Simulations. *Comput. Phys. Commun.* **2013**, *184*, 374−380.

(38) Valiev, M.; Bylaska, E. J.; Govind, N.; Kowalski, K.; Straatsma, T. P.; van Dam, H. J. J.; Wang, D.; Nieplocha, J.; Apra, E.; Windus, T. L.; et al. NWChem: A Comprehensive and Scalable Open-Source Solution for Large Scale Molecular Simulations. *Comput. Phys. Commun.* **2010**, *181*, 1477−1489.

(39) Fingerhut, B. P.; Costard, R.; Elsaesser, T. Predominance of Short Range Coulomb Forces in Phosphate-Water Interactions - a Theoretical Analysis. *J. Chem. Phys.* **2016**, *145*, 115101.

(40) Falk, M.; Hartman, K. A., Jr; Lord, R. C. Hydration of Deoxyribonucleic Acid. II. An Infrared Study. *J. Am. Chem. Soc.* **1963**, *85*, 387−391.

(41) Keller, P. B.; Hartman, K. A. The Effect of Ionic Environment and Mercury(II) Binding on the Alternative Structures of DNA. An Infrared Spectroscopic Study. *Spectrochim. Acta* **1986**, *42A*, 299−306.

(42) Shi, H.; Moore, P. B. The Crystal Structure of Yeast Phenylalanine tRNA at 1.93 Å Resolution: A Classical Structure Revisited. *RNA* **2000**, *6*, 1091−1105.

(43) Schauss, J.; Dahms, F.; Fingerhut, B. P.; Elsaesser, T. Phosphate-Magnesium Ion Interactions in Water Probed by Ultrafast Two-Dimensional Infrared Spectroscopy. *J. Phys. Chem. Lett.* **2019**, *10*, 238−243.

(44) Schauss, J.; Kundu, A.; Fingerhut, B. P.; Elsaesser, T. Contact Ion Pairs of Phosphate Groups in Water: Two-Dimensional Infrared Spectroscopy of Dimethyl Phosphate and Ab Initio Simulations. *J. Phys. Chem. Lett.* **2019**, *10*, 6281−6286.

(45) Mukamel, S. Multidimensional Femtosecond Correlation Spectroscopies of Electronic and Vibrational Excitations. *Annu. Rev. Phys. Chem.* **2000**, *51*, 691−729.

(46) Hamm, P.; Zanni, M. *Concepts and Methods of 2D Infrared Spectroscopy*; Cambridge University Press: Cambridge, 2011.

(47) Falk, M.; Hartman, K. A., Jr; Lord, R. C. Hydration of Deoxyribonucleic Acid. I. A Gravimetric Study. *J. Am. Chem. Soc.* **1962**, *84*, 3843−3846.